\documentclass{aa}
\usepackage{graphicx,psfig}
\usepackage{natbib}
\usepackage{txfonts}
\def\bp{\object{$\beta$\,Pictoris}}

\def\be{\begin{equation}}
\def\ee{\end{equation}}
\begin{document}
\title{Debris discs in binaries: a numerical study}

\author{P. Th\'ebault\inst{1}, F. Marzari\inst{2}, J.-C. Augereau\inst{3}}
\institute{
LESIA, Observatoire de Paris,
F-92195 Meudon Principal Cedex, France
\and
Department of Physics, University of Padova, Via Marzolo 8,
35131 Padova, Italy
\and 
Laboratoire d'Astrophysique de Grenoble (LAOG), Universit\'e Joseph Fourier,
B.P. 53, 38041 Grenoble Cedex 9, France}

\offprints{P. Th\'ebault} \mail{philippe.thebault@obspm.fr}
\date{Received; accepted} \titlerunning{Debris discs in binaries}
\authorrunning{Th\'ebault}

\abstract
%
{Debris disc analysis and modelling provide crucial information about the structure and the processes at play in extrasolar planetary systems. In binary systems, this issue is more complex because the disc should in addition respond to the companion star's perturbations.
}
%
{We explore the dynamical evolution of a collisionally active debris disc for different initial parent body populations, diverse binary configurations and optical depths. We focus on the radial extent and size distribution of the disc at a stationary state. 
}
{
We numerically follow the evolution of $10^{5}$ massless small grains, initially produced from a circumprimary disc of parent bodies following a size distribution in $dN \propto s^{-3.5}ds$ . Grains are submitted to both stars' gravity as well as radiation pressure. In addition, particles are assigned an empirically derived collisional lifetime.
}
{For all the binary configurations the disc extends far beyond the critical semimajor axis $a_{crit}$ for orbital stability. This is due to the steady production of small grains, placed on eccentric orbits reaching beyond $a_{crit}$ by radiation pressure. The amount of matter beyond $a_{crit}$ depends on the balance between collisional production and dynamical removal rates: it increases for more massive discs as well as for eccentric binaries. Another important effect is that, in the dynamically $stable$ region, the disc is depleted from its smallest grains. Both results could lead to observable signatures. 
}
{We have shown that a companion star can never fully truncate a collisionally active disc. For eccentric companions, grains in the unstable regions can significantly contribute to the thermal emission in the mid-IR. Discs with sharp outer edges, especially bright ones such as HR4796A, are probably shaped by other mechanisms. 

}
\maketitle

\section{Introduction} \label{intro}

\subsection{Planets and discs in binaries} \label{intro1}

Debris discs are dusty circumstellar systems observed around main sequence stars. They are thought to represent a late stage of disc evolution, when the primordial gas has been dispersed and most of the solids have been accreted onto embryos or fully formed planets \citep{lag00}. What is seen are small, typically $\leq 1$\,cm dust particles, produced by collisions among larger unseen parent bodies, probably leftover planetesimals not used in the planet building process \citep[see review by][]{wyatt08}. Most debris discs are initially detected through an excess IR flux in the stellar spectrum due to the thermal emission of dust. In some favourable cases resolved images are also obtained, in scattered light and, for some systems, also in thermal light at longer wavelengths. Most resolved discs do not appear homogeneous and smooth but show on the contrary pronounced spatial features, like clumps, warps, radial asymetries or spiral arms \citep{wyatt08}. Numerical and analytical models have been used to analyse these features and have provided precious information about the structure and the dynamics of these systems. In many cases, the presence of azimuthal, radial or vertical structures has been successfully explained as been due to the gravitaional pull of unseen perturbers, such as planets or planetary embryos. \footnote{The best known examples of model-infered perturbers are the Jovian planet producing the warp in the \bp\ disc \citep{moui97} and the one producing the inner edge of Fomalhaut's magnificent ring \citep{kalas05}. Both planets have since then been directly imaged \citep{kalas08,lag09}.}

In this respect, binaries are of special interest, since the secondary star's perturbations should significantly affect the extent, structure and evolution of a circumprimary debris disc and could induce pronounced features: confined rings, spiral arms, etc. More generally, studying dusty discs in binaries, especially young and bright discs that should be the observable counterparts of the last stages of planet formation \citep{ken04}, can provide crucial information about planet formation processes in these systems. Even fainter discs around older stars are of great interest since they should trace the location of asteroid or Kuiper belt-like belts, whose structure should provide clues to the presence of unseen planets. Comparing debris disc structures in binaries to that of single star systems could thus provide indirect but crucial information about how binarity affects planet formation processes.

Such information is very timely in the current context of a renewed interest for the problem of planet formation in binaries. Recent years have indeed seen a spectacular surge in the number of studies dedicated to this issue, fueled by the discovery of $\sim 80$ exoplanets in binaries, some of them in tight systems having a separation of less than 30\,AU \citep{desi07,mug09}. Some issues are today relatively well understood, for instance the limiting distance $a_{crit}$, as a function of the binary's configuration, for orbital stability around each star \citep[e.g][]{holw99,mud06}. This limit does also roughly correspond to the region where the last stages of planet formation, leading from lunar-sized embryos to fully formed planets, can proceed \citep{quin07}. However, some issues are still actively debated, in particular regarding the earlier stage leading from kilometre-sized planetesimals to the embryos \citep[e.g.][]{theb06,theb09,paard08,xie09,paard10}. 

As signposts of planet formation and/or planet location, debris discs are thus of crucial importance to understand what is going on in potentially planet forming binaries. However, to understand the connection between an observed dust population and the underlying architecture of the whole disc (collisional parent bodies, planets, etc.), it is crucial to first understand what imprint the binarity itself will leave on the disc.

\subsection{Previous studies} \label{previous}

Photometric surveys, looking at mid-IR excesses of debris discs around large stellar samples, have been attempting to identify trends specific to binary systems. The existence of such trends is however still debated. \citet{tril07} looked with {\it Spitzer} at 69 A- and F-type double stars and found that the disc frequency is lower for binaries with intermediate separation 3-30AU than for both tighter and wider binaries. On the contrary, \citet{plav09} and \citet{duch10} claim that no such trend is visible. Clearly, additional observations are needed to settle this issue. In any case, the theoretical part of \citet{tril07}'s study is of great interest. It argues that, since thermal emission at a given wavelength $\lambda_T$ probes a given dust temperature and thus a distance $a_{th}$ from a star of a given spectral type, then binaries having their $a_{crit} \leq a_{th} \leq a_{crit}^{cib}$ should have a lower excess rate at $\lambda_T$, because regions that are probed correspond to unstable orbits (where $a_{crit}^{cib}$ is the inner limit for stable {\it circumbinary} orbits).
In any case, the fact that binaries do have an effect on debris discs would seem a logical result in the light of recent studies that have shown that binaries affect the evolution of primordial protoplanetary discs in younger systems \citep{cieza09}.

Regarding spatially resolved systems, the most famous example of a debris disc in a binary is probably HR4796A, with a very bright ring at $\sim 70\,$AU from the central star and a physically bound companion at a projected distance of $\sim 510\,$AU \citep{jura93}. The possible impact of the companion on the disc has often been mentioned as a potentially important effect without having been investigated in details \citep{aug99,schnei09}. So far, the only system for which the effect of a companion star has been numerically investigated is HD 141569A. This A0V star is surrounded by a massive dusty and gaseous disk with a complex structure, including a spiral arm \citep{clam03}. It has two possibly bound M dwarf companions at projected distances of 750 and 890\,AU, whose influence on the dust disc has been studied in a couple of studies. \citet{aug04} have estimated the steady state reached by the disc after repeated passages of a companion of varying eccentricity $e_b$ and separation $a_b$, but neglecting the crucial effect of radiation pressure (see Sec.\ref{appro}). The same collisionless and radiation-pressure-less problem was investigated by \citet{quil05} using a hydrodynamical approach. \citet{ardi05} have considered the response of a dusty disc taking into account radiation pressure, but only for a single fly-by of a companion. More recently, \citet{reche09} also looked at the disc's response to a single fly-by.

\subsection{Present approach} \label{appro}

As can be seen, the debris-disc-in-a-binary problem has so far not been thoroughly investigated, i.e., only for one speficic system and each time only considering one specific aspect of the problem. We intend to take these studies a step further and consider this problem from a general perspective, exploring a large range of disc profiles as well as binary parameters (separation $a_b$ and eccentricity $e_b$). We will especially focus on one crucial effect, that of radiation pressure on the smaller grains. This mechanism might indeed greatly affect the response of the system to a companion star's perturbation. We can basically identify two potential effects:
\begin{itemize}
\item 1) Small grains are steadily produced by collisions from larger parent bodies, and are automatically placed on high-eccentricity orbits by radiation pressure. So that even if the parent bodies stay within the orbital stability region, high-$\beta$ grains might populate dynamically "forbidden" regions, depending on how fast they are produced as compared to how fast they are dynamically ejected.
\item 2) Because most small grains are on high-$e$ orbits with their apoastron potentially in unstable regions, they might be removed by the companion star even if they have been produced from parent bodies on stable orbits. In other words, companion star's perturbations could remove small grains from regions $within$ the stability limit.
\end{itemize} 
The purpose of this study is to quantitatively investigate these different effects. We use a specifically developed numerical model to follow the evolution of a collisional debris disc, orbiting a primary star and perturbed by a more distant companion. Grains are submitted to the radiation pressure of the primary star.

\section{Model} \label{model}

\subsection{Methodology} \label{methodo}

The system we intend to investigate is that of a debris disc orbiting a primary star and perturbed by a stellar companion. The dynamical evolution of the debris disc will be traced by a population of $N$ test particles, using a $4^{th}$ order Runge-Kutta integrator with adaptative step. We adopt the restricted 3 body problem, where particles are massless and only the potential of each star is considered. One crucial additional force is taken into account, i.e., radiation pressure from the central star.
With such a deterministic approach it is impossible to realistically follow the collisional evolution of the system, since the numerous fragments produced by each individual impact would lead to an exponential increase of $N$ rapidly exceeding any manageable number \footnote{a realistic treatment of collision outcomes would require a statistical  approach, with no or a simplified treatment of the dynamics \citep[e.g.][]{kriv06,theb07}}. However, collisions cannot be fully ignored for the present problem. A first reason is that they are imposing the {\it size-distribution} of the debris population. This parameter is here crucial since radiation pressure is a size-dependent force, whose global effect is related to the fraction of small, radiation-pressure affected grains. In addition, collisions set the rate at which grains are produced, a parameter that is in direct competition with the rate at which they can be removed by dynamical effects. 

In practice, we first run a simulation with $N_{PB}=10^{5}$ "parent bodies" not affected by radiation pressure ($\beta=0$) and wait until it reaches a steady state. By steady state, we mean that the spatial structure of the parent body disc is the same between two consecutive passages of the companion star at the same phase angle (the disc structure is the same at a time $t$ and at time $t+t_{bin}$). We then consider a set of 10 parent body disc configurations at 10 regularly spaced time intervals within one binary orbital period. Following a procedure similar to \citet{ardi05}, we generate, from each  parent body configuration, a dust population following a collisional equilibrium size distribution \citep{dohn69} in $dN \propto s^{-3.5}ds$ (i.e., $dN \propto \beta^{1.5}d\beta$), from $\beta=0.012$ down to very small unbound grains with $\beta=2.5$, and we compute its evolution in time \footnote{Note that the size distribution in real systems with a minimum size cut-off should depart from this power-law and exhibit a characteristic wavy profile \citep[e.g][]{camp94,theb07}. However, we shall ignore these refinements here and simply assume the Dohnanyi profile}. 
Dust particles are progressively removed either by radiation pressure for grains with $\beta>0.5$, or by collisional destruction or by dynamical perturbations from the companion star. For the dynamical ejection, we remove particles when they pass at less than 0.1 Hill radius of the companion, or when they are on an unbound orbit $and$ at a radial distance exceeding $50a_{crit}$. As mentioned earlier, collisions cannot be properly taken into account with our deterministic code, so we consider a simplified approach in which, during each timestep $dt$ spent \emph{within the collisionally active parent body region}, each particle is assigned a collisional destruction probability
\begin{equation}
f_{Dcoll} = \frac{dt}{t_{Dcoll}} = \frac{4\pi\tau}{t_{orb}}\,dt
\label{fcoll}
\end{equation}
where $t_{Dcoll}$ is a characteristic collisional destruction timescale. We assume that most collisions are destructive, so that $t_{Dcoll}$ is comparable to the collision timescale $t_{coll}$  for which we follow \citet{theb05} and assume
\begin{equation}
t_{coll} = \left(\frac{s}{s_0}\right)^{\alpha} \frac{\delta V}{\delta V_0}\,\,\,t_{coll_0}
\label{fcoll2}
\end{equation}
where $s_0$ is a reference size, $\delta V$ and $\delta V_0$ are the departures from the local Keplerian velocity $V_{Kep}$ of the considered particle and for a parent body with $\beta = 0$ respectively, and $t_{coll_0}=t_{orb}/(4\pi\,\tau)$ is the classical size-independent simplified expression, proportional to the local orbital period $t_{orb}$ divided by the local total cross sectional area, or vertical optical depth $\tau$. \footnote{The apparent difference with \citet{theb05}'s expression comes from the fact that the $t_{coll_0}$ considered here implicitly regroups several parameters of \citet{theb05}'s Equ.6}. For $s_0$ we consider the size of the smallest particles that behave like parent bodies, assumed to be the size for which the eccentricity imposed by radiation pressure $e_{\beta} = \beta/(1.-\beta)$ is smaller than the eccentricity assumed for the parent bodies $e_0$. For the size dependence index $\alpha$, we assume the value -0.3, as obtained by \citet{wyatt02}.
Note that the collision rate is assumed to be independent of the companion star's orbit and of the dynamical perturbations it triggers in the birth ring. This simplification is acceptable as long as the particle-in-a-box approximation for estimating encounter rates holds. However, this approximation breaks down for particles on high-$e$ orbits \citep{wyatt10}, which could be the case for the highest $e_b$ considered here, i.e. 0.75, for which orbits in the parent body disc might reach values as high as $\sim 0.3$. Such values correspond to the limit where a more refined model like that of \citet{wyatt10} is required. Although implementing such a model is beyond the scope of the present work, we would thus like to point to possible limitations for our $t_{coll}$ estimate in the high $e_b$ cases (for more discussion on the collision rate approximation, see also Sec.\ref{sized}).

We then follow the same procedure as in \citet{theb05} and \citet{kriv09} and record, at {\it regularly spaced} time intervals, the positions of all remaining particles. The disc's spatial structure is progressively obtained by adding up these instantaneous snapshots until all particles have been removed either dynamically or collisionally. This procedure is implicitly equivalent to assuming an arbitrarily constant dust production rate from parent bodies \citep[see discussion in][]{theb05}. 
By this method we obtain 10 disc profiles corresponding to our sample of 10 different initial parent body configurations. The final disc profile is obtained by averaging these different profiles.

\subsection{Numerical set up} \label{setup}
\begin{table}
\begin{minipage}{\columnwidth}
\caption[]{Set up}
\renewcommand{\footnoterule}{}
\label{init}
\begin{tabular*}{\columnwidth} {ll}
\hline
Radial extent \footnote{normalized to $a_{crit}$} & $0.6<a<1$ (extended disc case)\\
 &  $0.55<a<0.6$ (inner ring case) \\
Eccentricity of the parent bodies $e_0$ & 0.05\\
Surface density profile (parent bodies)& $\Sigma =$ cst\\
Disc optical depth for $t_{coll}$ estimation & $\tau = 2\times10^{-3}$ (dense disc)\\
 & $\tau = 2\times10^{-4}$ (faint disc)\\
Number of test particles $N$ & $10^{5}$\\
Size range \footnote{as parameterized by $\beta \propto 1/s$} & $\beta(s_{max})=0.012\leq\beta(s)\leq\beta(s_{min})=2.5$\\
Size distribution at $t=0$ & $dN(s) \propto s^{-3.5}ds$\\
\hline
\end{tabular*}
\end{minipage}
\end{table}

For the sake of clarity and so that the results can be easily understood in terms of departures from the purely gravitational case, all distances will be scaled to $a_{crit}$, the estimated outer limit for orbital stability corresponding to a given binary configuration $a_b,e_b,\mu$ in the purely gravitational case (i.e., when negecting radiation pressure, see Sec.\ref{intro1}). We compute $a_{crit}$ using the empirical formulae by \citet{holw99}
\begin{equation}
\frac{a_{crit}}{a_b} = 0.464\\
-  0.380\mu - 0.631e_b + 0.586\mu e_b + 0.150e_{b}^{2} - 0.198\mu e_{b}^{2}
\label{hw}
\end{equation}
where $\mu=m_2/(m_1+m_2)$, with $m_1$ the mass of the primary and $m_2$ that of the companion star.
Note that there exists other estimates of this parameter, but the \citet{holw99} estimate has proven to be relatively accurate for most cases and is thus satisfying as a first order approximation of the width of the dynamically stable circumprimary region. In any case, let us stress that our results do not depend on the assumed value for $a_{crit}$, since we automatically obtain our stable parent body disc from specific test runs \footnote{the accuracy of the \citet{holw99} estimate is confirmed by the fact that, for all our parent body runs, the maximal extent of the parent body disc is in good agreement with $a_{crit}$}. The $a_{crit}$ parameter is just used as a convenient, and dynamically relevant scaling factor.

Given the number of free parameters, a thorough phase space exploration is impossible. We cannot in particular explore all possible disc and binary configurations. We will thus restrict ourselves to two disc geometries: 1) A nominal "wide disc" case where the parent body extends up to $a_{crit}$, 2) an "inner ring" case, whose outer edge lies far within the orbital stability limit. 
For the disc's optical depth, we also consider two cases: 1) a nominal case with a dense and collisionally active disc with $\tau = 2\times10^{-3}$ corresponding to bright debris discs such as HD32297 or \bp\ (the brightest systems, such as HR4796 have even higher $\tau$, see Sec.\ref{hrfit}), and 2) a low $\tau = 2\times10^{-4}$ case modelling fainter systems, with lower collision rates, such as AUMic.
For each disc case, a full set of binary configuration cases will be run. We explore a wide range of $e_b$ values, ranging from 0 to 0.75. For each $e_b$, the value of $a_b$ is then automatically given by the condition that $a_{crit}$ is kept constant. The only parameter that will not be thoroughly explored is the mass ratio, for which we assume a fixed value $\mu=0.25$ (except for the specific HR4796A run, see Sec.\ref{hrfit}), which corresponds to estimates of the mean mass ratio in binaries derived from extensive surveys \citep{kroupa90,duq91}.

For the dust population, we consider $N = 10^{5}$ test particles, launched from the parent body ring following a $dN \propto s^{-3.5}ds$ size distribution. The particles' absolute sizes are of no importance here, the crucial parameter being their $\beta$ value. As a consequence, sizes will be scaled by this $\beta$ parameter, and we take $\beta_{min}=0.012$ (maximum size) and $\beta_{max}=2.5$. Due to the steep size distribution, we run in practice 3 separate runs with $N=10^{5}$ particles, for the $0.0012\leq\beta\leq 0.083$, $0.083\leq\beta\leq 0.5$ and $0.083\leq\beta\leq 2.5$ domains respectively, which are then weighted and recombined.

All initial conditions are summarized in Tab.\ref{init}.

\section{Results} \label{res}

\subsection{Test run for single star}

\begin{figure}
\includegraphics[angle=0,origin=br,width=\columnwidth]{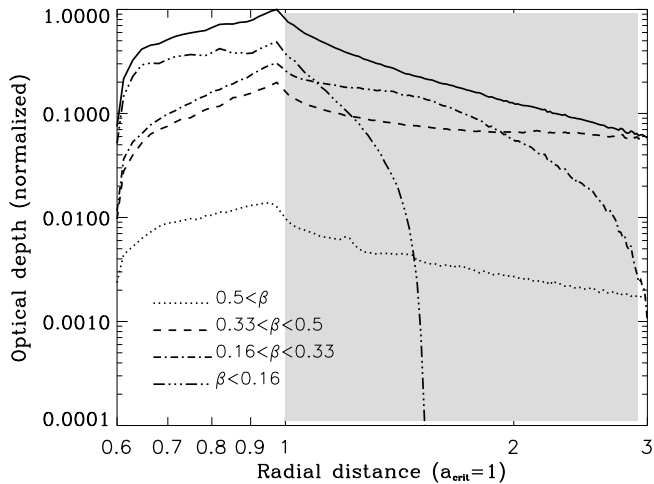}
\caption[]{Run with no companion star. Normalized and azimutally averaged radial profile of the vertical optical depth for the whole grain population (solid line), and for three sub-populations corresponding to different grain size (i.e., $\beta$) range. The initial parent body disc ($\beta=0$ particles) extends to $a_{crit}=1$. }
\label{rwnob}
\end{figure}

We first run a test simulation without a binary companion, i.e., just consisting of an initial disc of parent bodies extending to $a_{crit}=1$ ($a_{crit}$ is here only a reference for comparison with the cases with companion where it has a physical meaning). Its only evolution is thus due to the analytical collisional removal prescription. Fig.\ref{rwnob} shows the steady state reached by the system. Not surprisingly, the region beyond $a_{crit}$ is populated by small, high-$\beta$ grains launched on high eccentricity orbits because of radiation pressure. The optical depth profile beyond the parent body region has a radial dependence that follows a power law in $\propto r^{-1.5}$. This is a well known result for collisional systems at steady state, for which it has been shown that the vertical optical depth naturally assumes a $r^{-1.5}$ profile ($r^{-3.5}$ for the mid-plane luminosity) beyond the collisionally active region of parent bodies \citep{stru06,theb08}. 
Note that the contribution of unbound grains ($\beta>0.5$) is negligible in the whole considered radial region (see Sec.\ref{sized} a more detailed discussion).

\subsection{Perturbed disc structure in the presence of a companion star}

\begin{figure}
\includegraphics[angle=0,origin=br,width=\columnwidth]{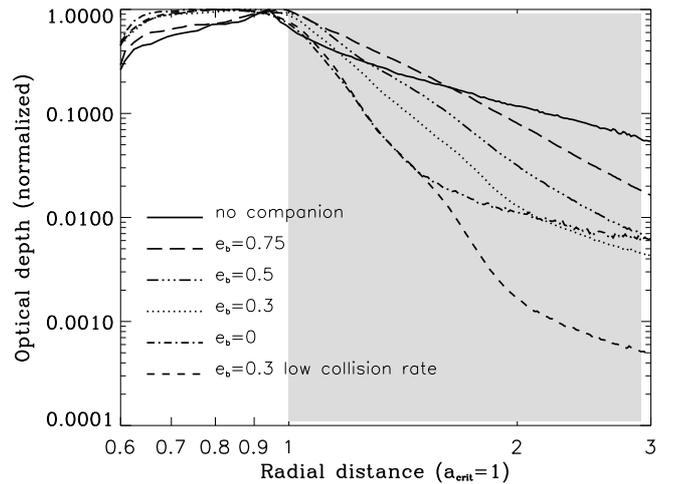}
\caption[]{Azimutally averaged radial profile of the vertical optical depth for 6 different cases. All profiles have been normalized to their value at $a=a_{crit}$.)}
\label{radwide}
\end{figure}

\begin{figure}
\includegraphics[angle=0,origin=br,width=\columnwidth]{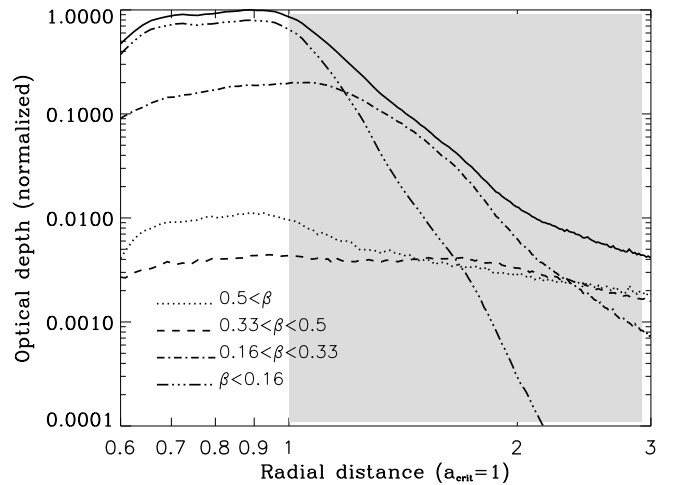}
\caption[]{Same as Fig.\ref{rwnob} but with a companion star with $e_b=0.3$ such as the dynamical stability limit around the primary is at $a_{crit}=1$.}
\label{rwbin}
\end{figure}
\begin{figure*}
\makebox[\textwidth]{
\includegraphics[width=\columnwidth]{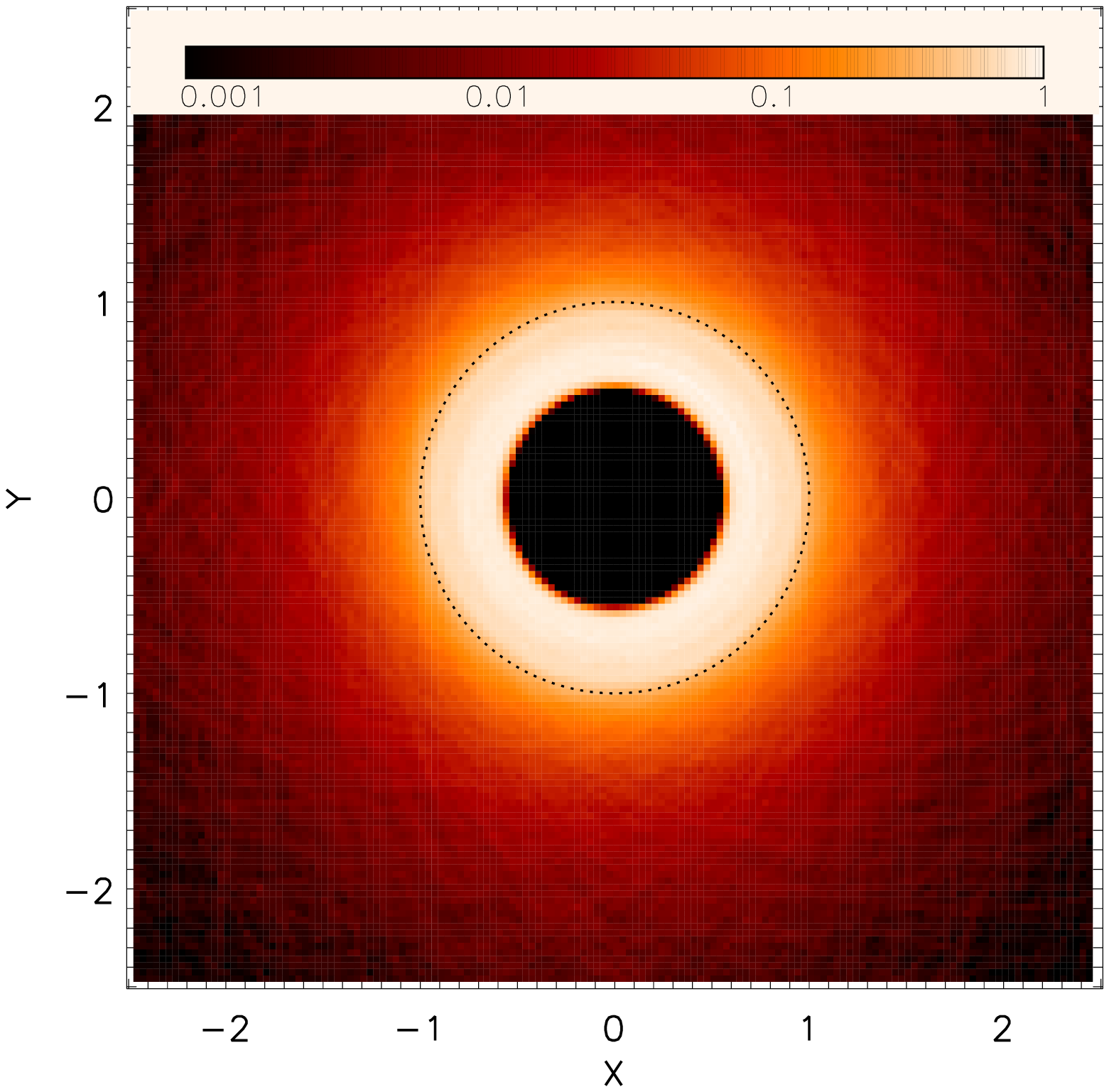}
\hfil
\includegraphics[width=\columnwidth]{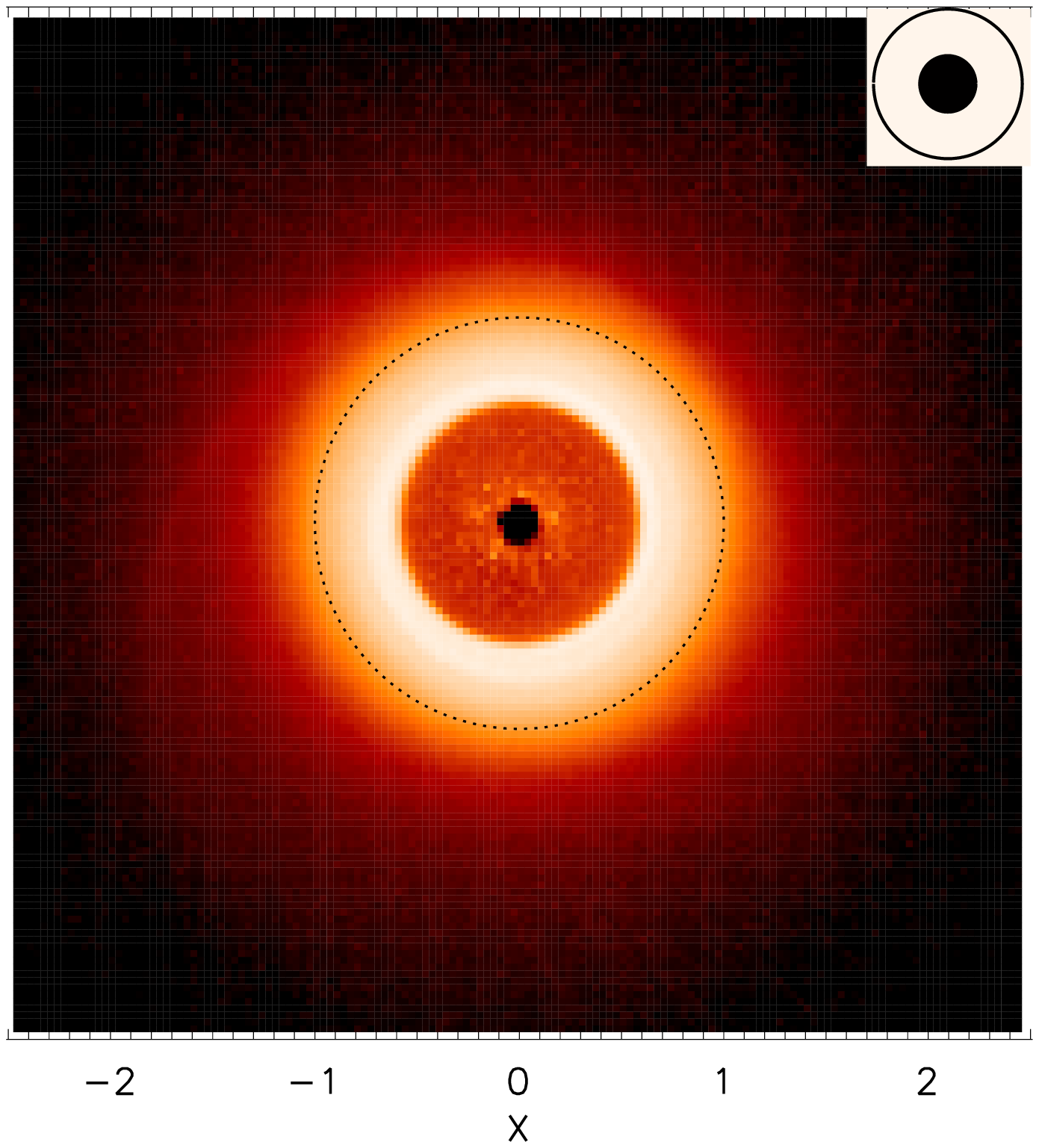}
}

\makebox[\textwidth]{
\includegraphics[width=\columnwidth]{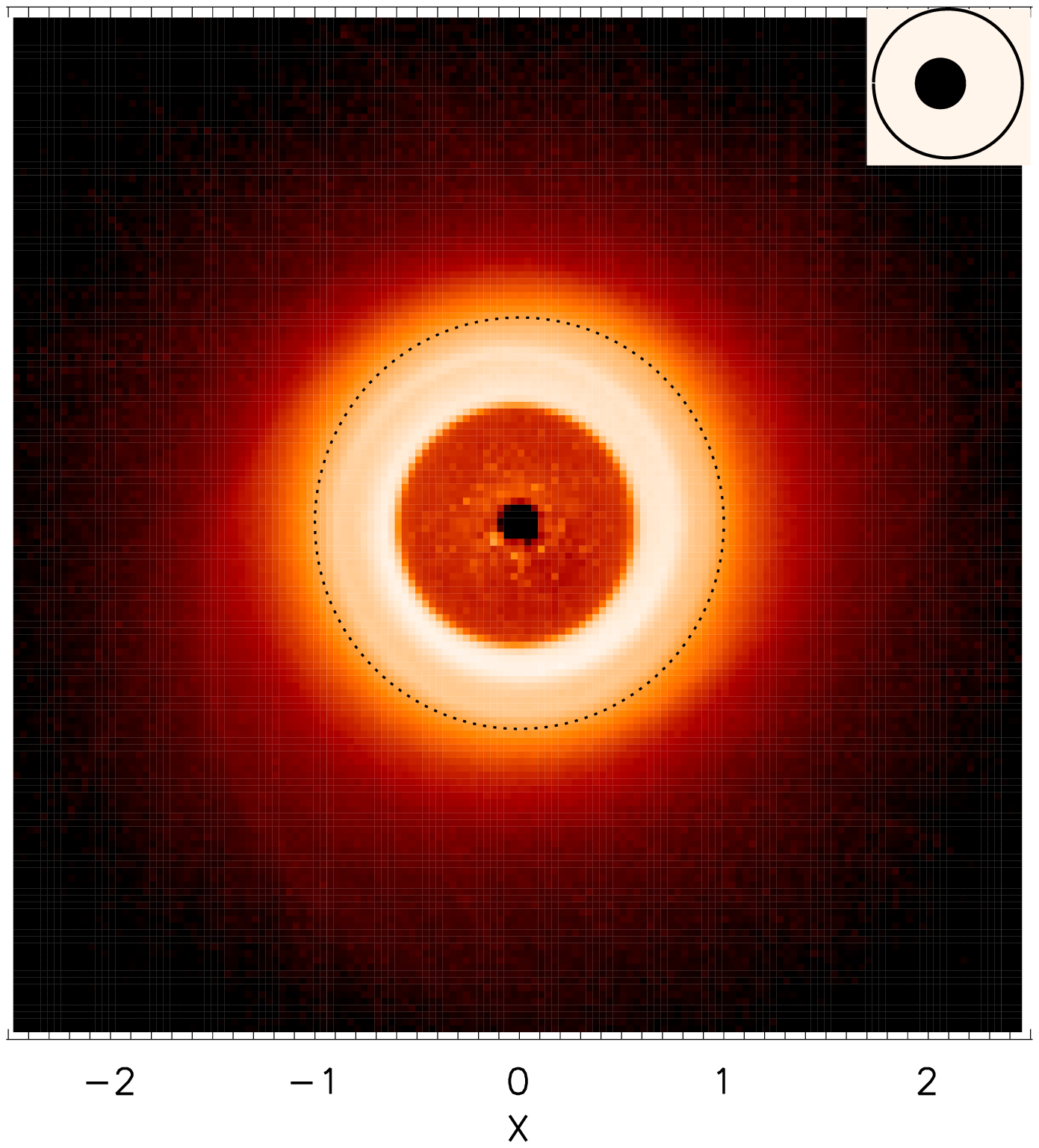}
\hfil
\includegraphics[width=\columnwidth]{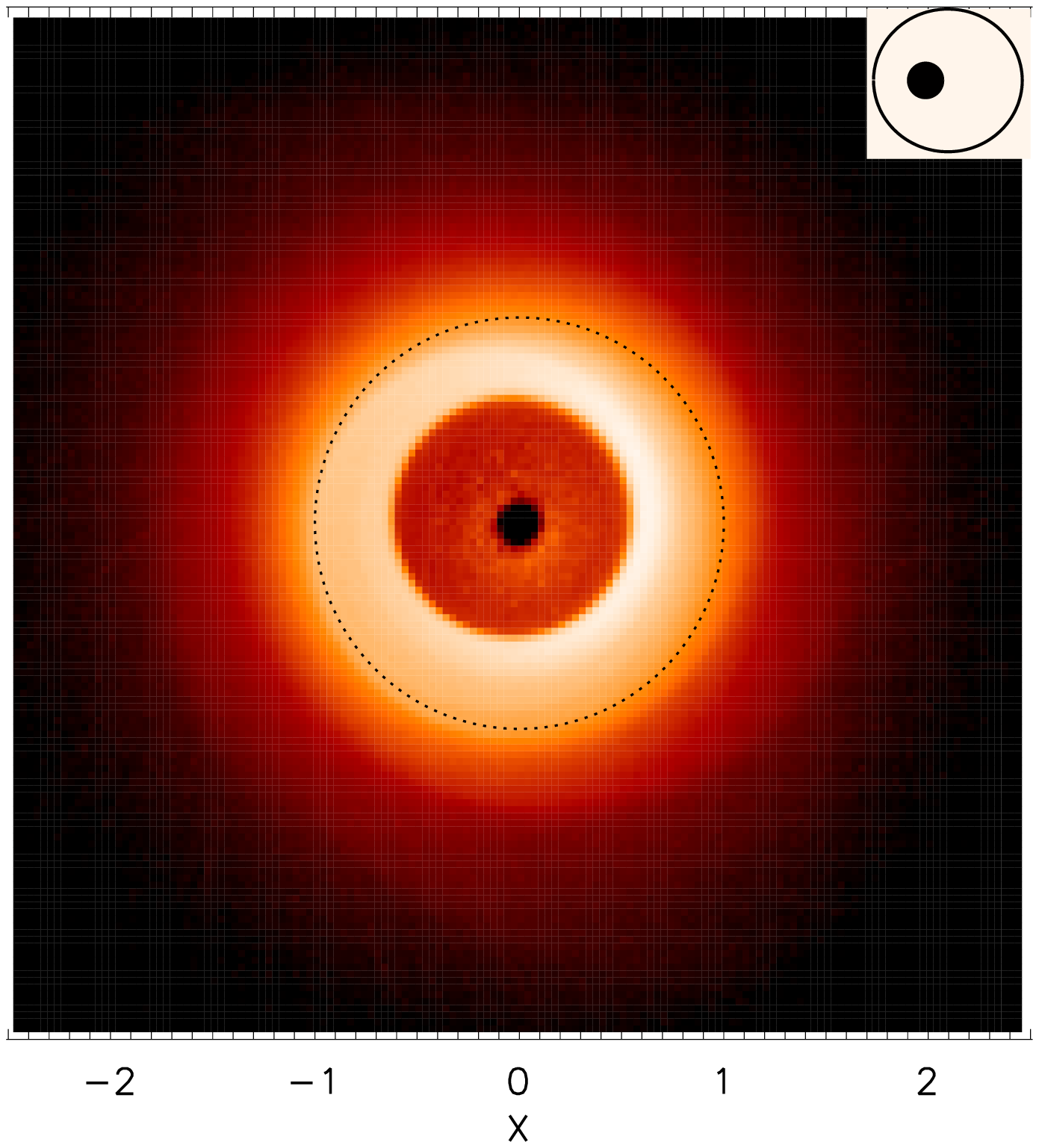}
}
\makebox[\textwidth]{
\includegraphics[width=\columnwidth]{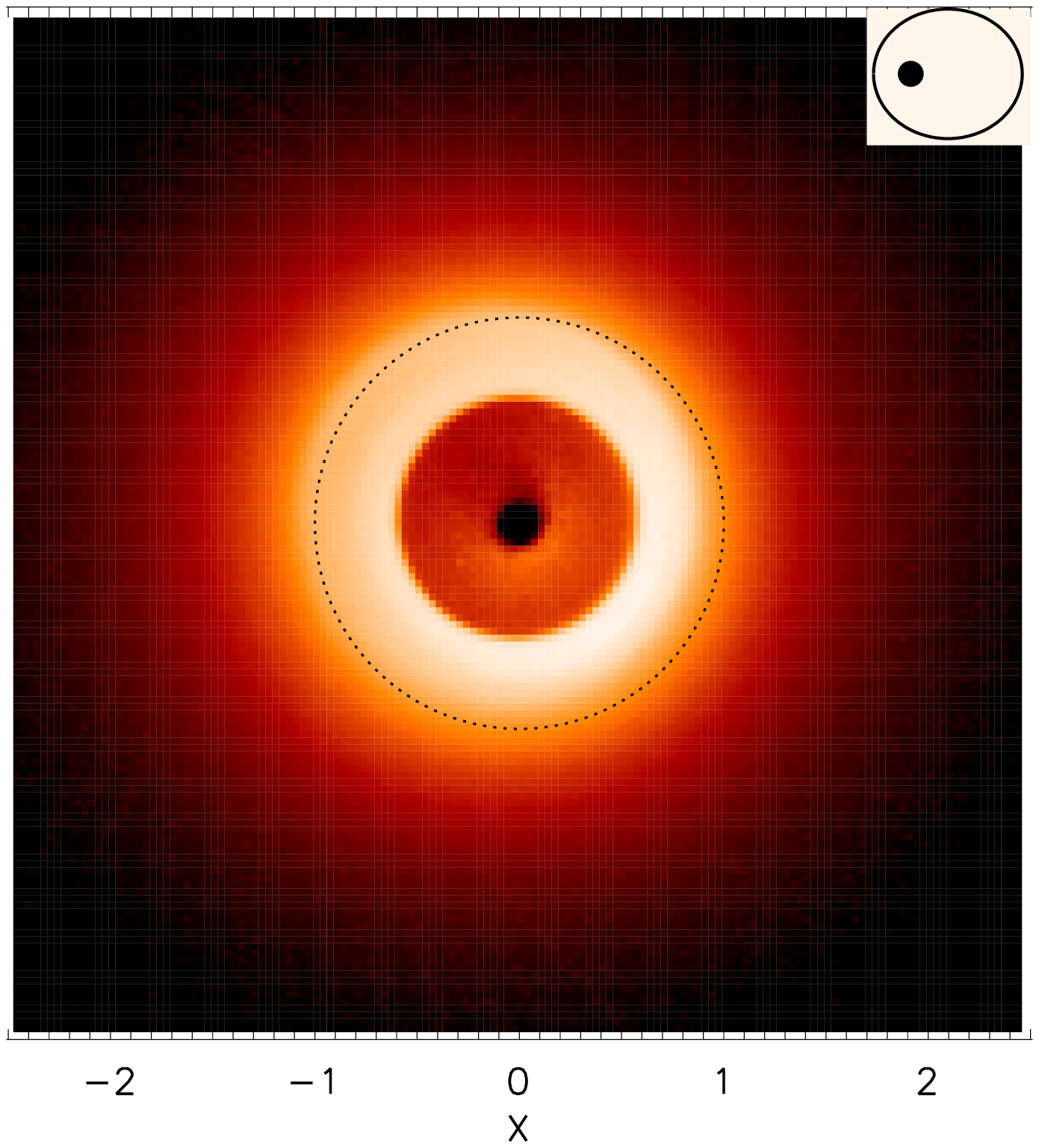}
\hfil
\includegraphics[width=\columnwidth]{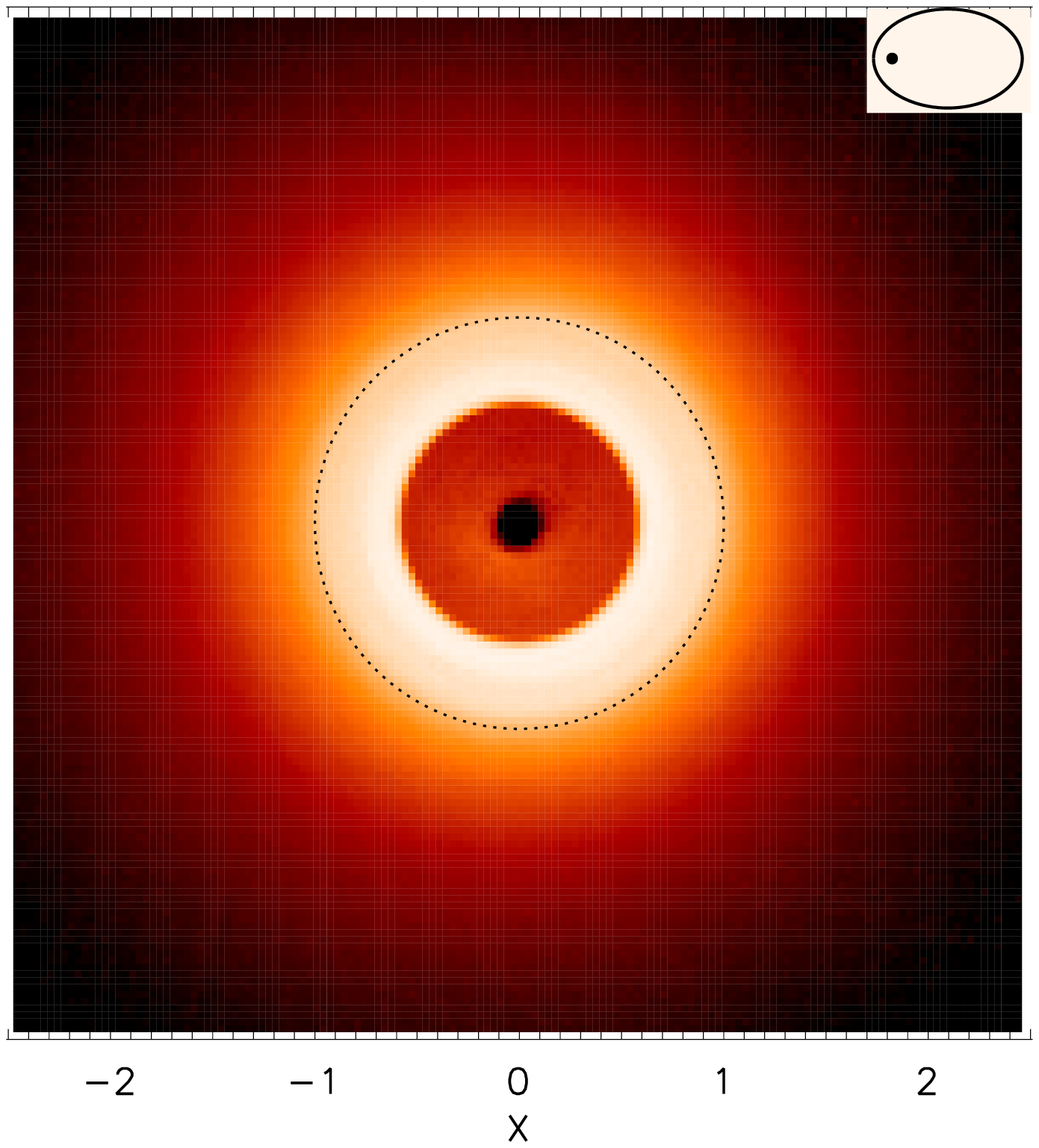}
}
\caption[]{ Normalized synthetic images in scattered light, at steady state, for a wide disc of parent bodies initially extending to $a_{crit}$, and for different orbital configuration of the binary, where the value $a_{crit}$ is kept constant and $e_b$ varies. The upper left panel is a test case with no companion and the next 5 runs are for $e_b=$0, 0.1, 0.3, 0.5 and 0.75 respectively. The location of the $a_{crit}$ radius is shown by the dashed black circle. The binary configuration is schematized in the small white square at upper right of each panel, where the region within $a_{crit}$ is shown as a filled black circle.
All distances have been normalized so that $a_{crit}=1$.
}
\label{snapsh}
\end{figure*}

In the following simulations we include the companion star and explore several binary configurations, by varying $e_b$ and $a_b$ while keeping $a_{crit}=1$ as a reference anchor point. The optical depth $\tau$ for the disc is for most cases that of our nominal dense disc (high collision rate) case, but the faint disc $\tau=2\times10^{-4}$ case is also explored. The main outcomes are summarized in Fig.\ref{radwide}, which displays the averaged radial profile of the vertical optical depth, at steady state, for all cases.

The main result is that, for all cases, the dusty disc always extends well beyond the purely dynamical orbital stability radius $a_{crit}$. This is due to the steady production of small, high-$\beta$ grains that are produced within $a_{crit}$ but are launched on eccentric orbits reaching far beyond this limit. The dust population in the dynamically unstable region is larger and extends farther out for higher eccentricities of the binary. In effect, {\it for a fixed $a_{crit}$}, a companion on a circular orbit is the most effective in truncating the circumprimary disc, while eccentric companions have a more limited influence. This is because perturbers on eccentric orbits spend a large fraction close to their apoastron far away from the primary, thus allowing small grains that are produced in the circumprimay disc to move towards the outer parts of their (radiation pressure induced) eccentric orbits without encountering the secondary star. True, most of these grains will eventually be removed by the companion's perturbations, but this removal will take more time than for a circular orbit companion that always moves in the vicinity of the primary. This increased removal time allows for more grains to be present at a given time, since the collisional production time is, to a first order, independent of the binary configuration.
In addition, the contribution of unbound grains ($\beta>0.5$), which is negligible in the no-companion case, becomes significant in the $r>a_{crit}$ regions, as can clearly be seen in Fig.\ref{rwbin}. This is because these grains, contrary to all the others, do not get further depleted by the presence of the companion star (see also Sec.\ref{sized}). 

Of course, even for the highest $e_b$ cases, there is always less dust beyond $a_{crit}$ than for the companion free case, as is clearly illustrated by Fig.\ref{radwide}, showing the averaged radial profile of the vertical optical depth. Taking as the reference the ratio $\tau_{(2a_{crit})}/\tau_{(a_{crit})}$, we see that, for the $e_b=0$ case, it can be a factor 10  lower than for the companion free profile where $\tau_{(2a_{crit})}/\tau_{(a_{crit})} \simeq 0.1$. For the more eccentric binary cases, however, the difference with the no-companion case is much more limited. It is only a factor $\simeq\,3$ below the companion-free case for the $e_b=0.5$ run and less than a factor 1.5 for $e_b=0.75$. 
Note that, for all cases, the optical depth profile within the parent body disc slightly increases from $0.6a_{crit}$ to $a_{crit}$. This is due to the contribution of particles with high $\beta$ that are more numerous at the outer edge of the ring, where a large fraction of high-$\beta$ grains produced further in can be observed, than at the inner edge, where only locally produced high-$\beta$ grains contribute. When only considering the contribution of large particles with low $\beta$, we observe a flat optical depth profile similar to that for the input parent bodies (see Figs.\ref{rwnob} and \ref{rwbin}). A logical consequence is that profiles in the birth ring for efficiently truncated discs (those with low $e_b$ companions) are flatter than for less efficently truncated ones, because in the former cases the birth ring is more strongly depleted of small grains (and thus more dominated by low $\beta$ particles) than in the latter ones (see Sec.\ref{sized}).

For illustrative purposes, we show in Fig.\ref{snapsh} synthetic head-on seen images, in scattered light assuming gray optical scattering for the grains, for a set of runs with $e_b$ ranging from 0 to 0.75, as well as for the no-companion case for comparison. Note that these "images" cannot reproduce fine spatial structures, such as spiral arms or sharp azimuthal asymmetries, because of the phase averaging procedure used in producing the steady-state distribution. However, they clearly show that resolved observations with a sensitivity range of $10^{2}$ could reveal flux coming from dynamically forbidden regions.

\subsection{Size distribution in the parent body region} \label{sized}

\begin{figure}
\includegraphics[angle=0,origin=br,width=\columnwidth]{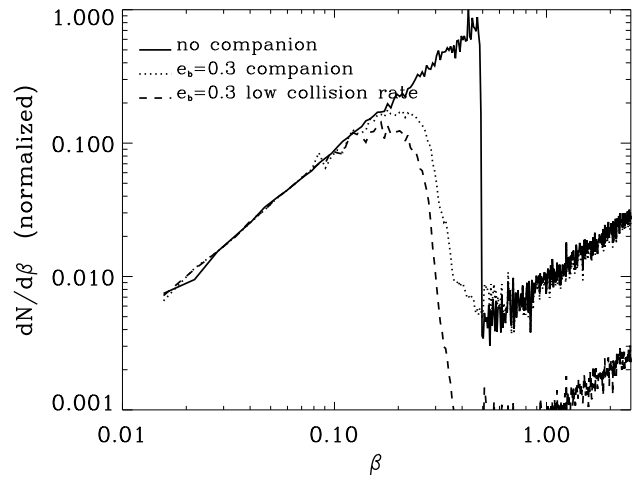}
\caption[]{Differential $\beta(s)$ (i.e., size) distribution {\it within the parent body disc} for the no-companion and the $e_b=0.3$ companion cases, as well as for the same $e_b=0.3$ case with a lower collision rate ($\tau = 2\times 10^{-4}$)}
\label{betwide}
\end{figure}

Fig.\ref{betwide} shows the steady state size distribution {\it within the parent body region} for both the non-companion and the reference $e_b=0.3$ cases. For the companion-free case, the differential size distribution has some small wiggles in the $0.2\leq \beta \leq 0.5$ domain but roughly follows a power law in $dN\propto \beta^{1.6}d\beta$, corresponding to $dN\propto s^{-3.6}ds$, which is relatively close to the input size distribution. More importantly, we do not observe a depletion of small, high-$\beta$ grains within the birth ring, despite of the fact that these small grains are on high eccentricity orbits and spend most of their time $outside$ of the ring.
This confirms the results of \citet{stru06} and \citet{theb08} that debris discs tend toward a state where the collisional equilibrium distribution (be it in $s^{-3.5}$ or not) holds within the collisionally active region, meaning that for the {\it system integrated} distribution, small high-$\beta$ grains are in excess.
Another expected result is the sharp transition at the blow-out size ($\beta=0.5)$, with a strong depletion of unbound grains. This is because these grains leave the system on a dynamical timescale while bound particles have a lifetime imposed by collisional effects. For the considered value of the system's optical depth, a few $10^{-3}$, the former is much shorter than the latter.

The effect of the companion on the size distribution is to deplete it from its smallest particles. For the $e_b=0.3$ case, this depletion becomes significant for $\beta \geq 0.15$ grains, while $\beta \geq 0.25$ particles are almost completely absent. Note that this depletion occurs in the parent body disc, {\it within} the orbital stability limit $a_{crit}$. It is due to the fact that small particles produced at $r<a_{crit}$ are placed by radiation pressure on eccentric orbits that enter the potentially unstable region beyond that limit. Depending on their position relative to the companion while in the $r<a_{crit}$ region, some of these small grains will be ejected and will not come back at their periastron in the parent body ring. As could be logically expected, this small grain depletion gets more pronounced when decreasing the disc's optical depth ($\tau = 2\times10^{-4}$ case), because the collision production rate of small grains is lower, while the removal timescale, only due to the dynamics, remains the same.
Another noteworthy result is that, for the same given optical depth, the number of unbound grains is approximately the same with or without the secondary star. This is a logical result in itself, because the removal rate of these small grains, imposed by radiation pressure blow out, is only marginally affected by the presence of the companion. 
An interesting consequence is that there is no longer a sharp transition at $\beta=0.5$. The excess of grains just below the $\beta=0.5$ threshold has been erased by the fact that bound particles in the $0.15\leq\beta\leq0.5$ range are strongly depleted by the secondary star. This is why the contribution of unbound grains to the total optical depth becomes non negligible, as has been seen in Fig.\ref{radwide}.

Note that we do not observe here the "wavy" size distribution, starting at the blow-out size ($\beta\sim 0.5$) imposed by radiation pressure, expected in realistic collisional systems \citep{theb07}. This is because our expression for $t_{coll}$ is too simplified to account for such complex effects. In order to do so, we could in principle use the empirical expression derived in Equ.7 of \citet{theb07}. Nevertheless, this expression is only valid for $isolated$ debris discs where the collisional cascade is not affected by dynamical perturbers \footnote{When using the expression of \citet{theb07} instead of our Equ.\ref{fcoll2} in the non perturbed single-star case we do indeed obtain a pronounced wavy pattern for the size distribution}.
This is clearly not the case here, because the companion star strongly modifies the abundances of grains close to the blow-out size. It tends in particular to erase the discontinuity at the $\beta=0.5$ limit that is the very source of the wavy distribution. We would thus expect a much less pronounced oscillation in the size distribution, although a secondary wave could possibly start at the limiting size for the depletion by the binary ($\beta \sim 0.15$). Studying these effects would require to use a code that can self-consistently follow the coupled dynamical and collisional evolution of the system. Such a task exceeds by far the scope of this paper, as there is to our knowledge no such all-encompassing code available yet.
Fig.\ref{betwide} should thus be regarded as a first approximation. However, to check the robustness of our results we performed a test run using the $t_{coll}$ expression of \citet{theb07} for the $e_b=0.3$ companion case. We verify that, even for this highly unrealistic case artificially forcing an effect at the $\beta=0.5$ limit, the resulting size distribution is remarkably similar to that of Fig.\ref{betwide}. The main part of the wavyness being completely erased by the dynamical depletion in the $\beta \sim 0.15$ domain. We are thus relatively confident that our main result, i.e. the binary-induced cut-off at $\beta \sim 0.15$, should be relatively robust given its amplitude, which exceeds by far that of any wavyness observed in detailed collisional studies \citep[see for instance][]{kriv06,theb07}.

\subsection{Inner ring case} \label{inner}

\begin{figure}
\includegraphics[angle=0,origin=br,width=\columnwidth]{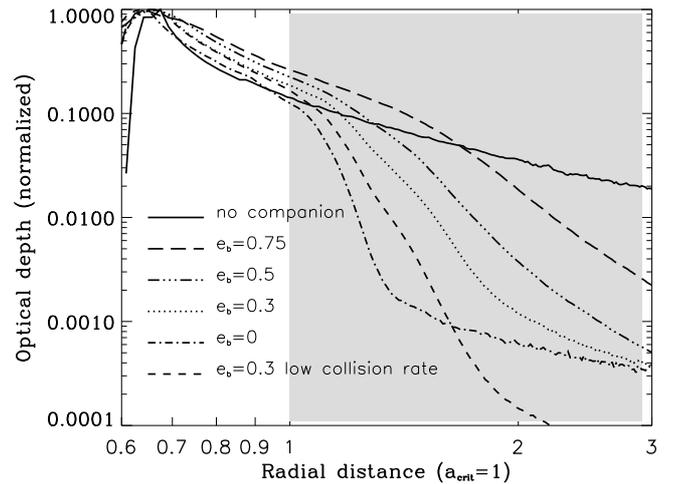}
\caption[]{Same as Fig.\ref{radwide} but for a parent body ring extending only to $0.6\,a_{crit}$.}
\label{radna}
\end{figure}
\begin{figure}
\includegraphics[angle=0,origin=br,width=\columnwidth]{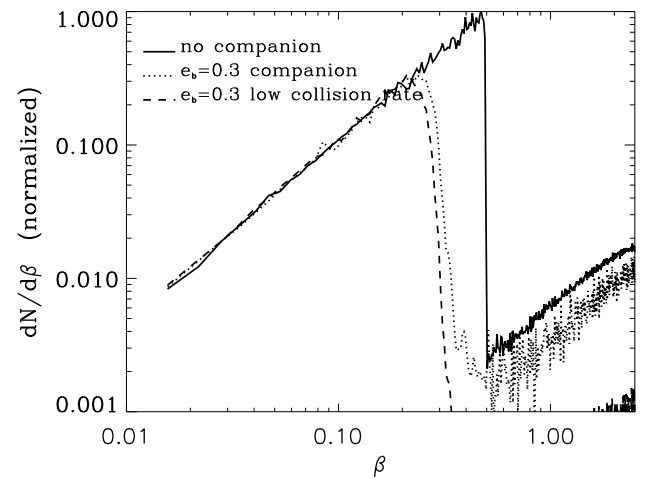}
\caption[]{Same as Fig.\ref{betwide}, but for the inner ring case.}
\label{betna}
\end{figure}
We consider here a parent body ring extending only to $a_{out}=0.6a_{crit}$. This corresponds in practice to astrophysical cases with greater binary separation, where the companion star didn't truncate the initial protoplanetary ring and $a_{out}$ is the natural disc outer limit. As it can be seen in Fig.\ref{radna}, even in this case, the debris disc extends well beyond $a_{crit}$ for all explored cases. However, the amount of material in the unstable zone is smaller than in the case where the disc extends all the way to $a_{crit}$. Taking again as a reference the ratio $\tau_{(2a_{crit})}/\tau_{(a_{crit})}$, we see that it is one order of magnitude lower than in the unperturbed case (no companion star) for binaries with $e_b\leq0.5$. For systems with higher eccentricites, however, the difference with the binary-free case drops to less than 50\%, a value comparable to that of the previous case with an extended disc. 
As for the size distribution within the parent body ring (Fig.\ref{betna}), results are remarkably similar to those with $a_{out}=a_{crit}$. It is depleted from most high-$\beta$ grains, the limiting $\beta$ value being only slightly higher, $\sim 0.2$ instead of $0.15$.

\section{Discussion} \label{Discussion}

\begin{figure*}
\makebox[\textwidth]{
\includegraphics[width=\columnwidth]{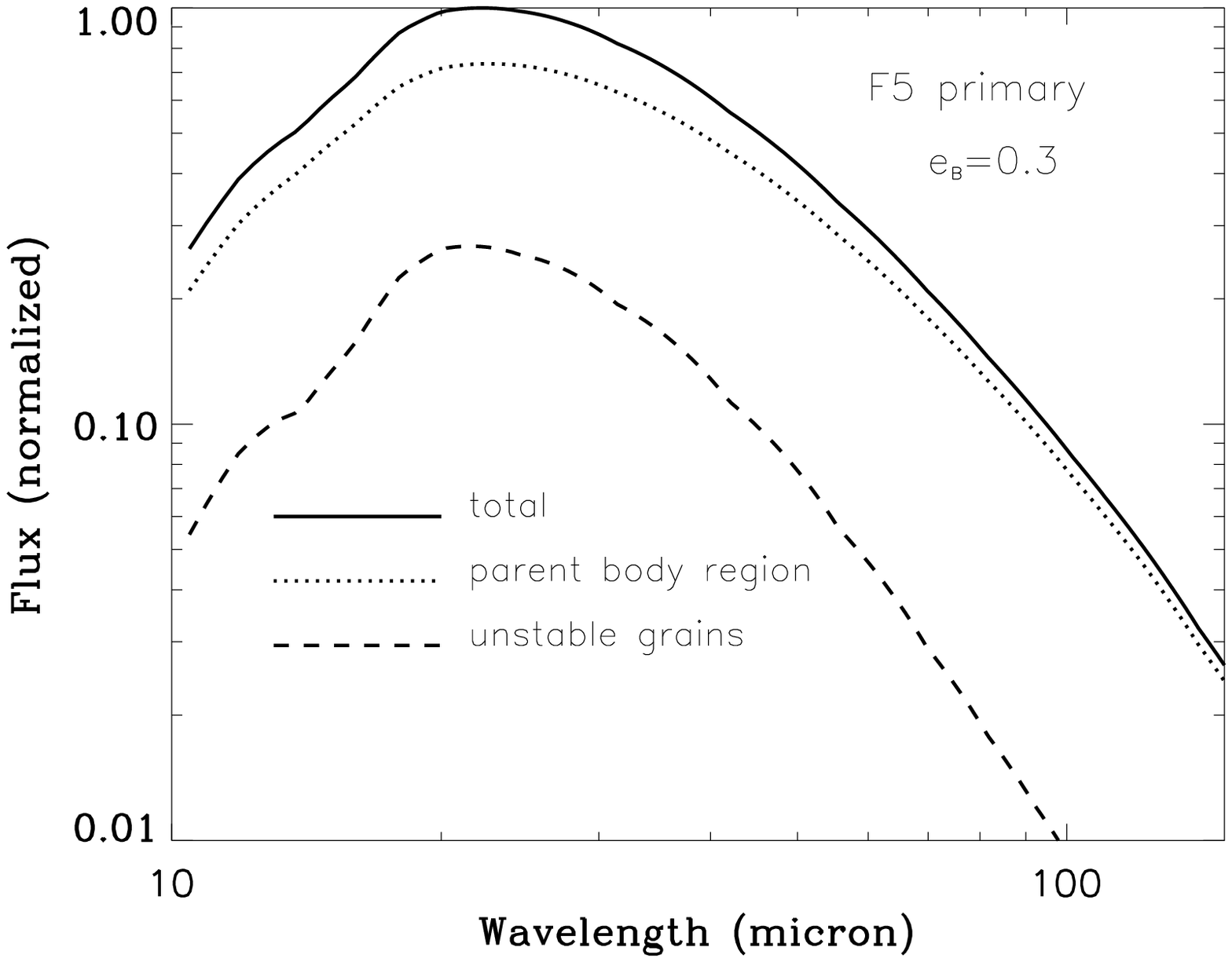}
\hfil
\includegraphics[width=\columnwidth]{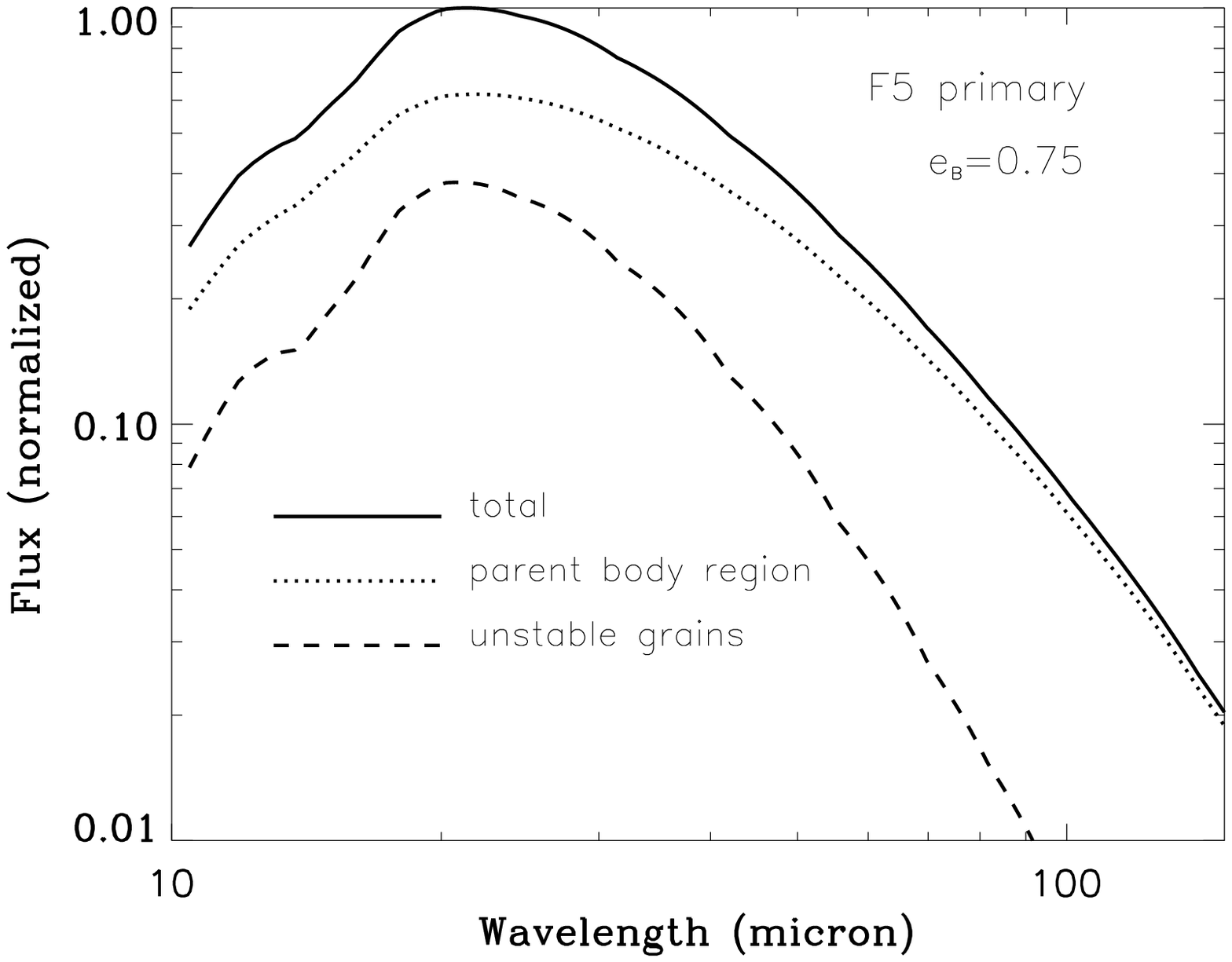}
}

\makebox[\textwidth]{
\includegraphics[width=\columnwidth]{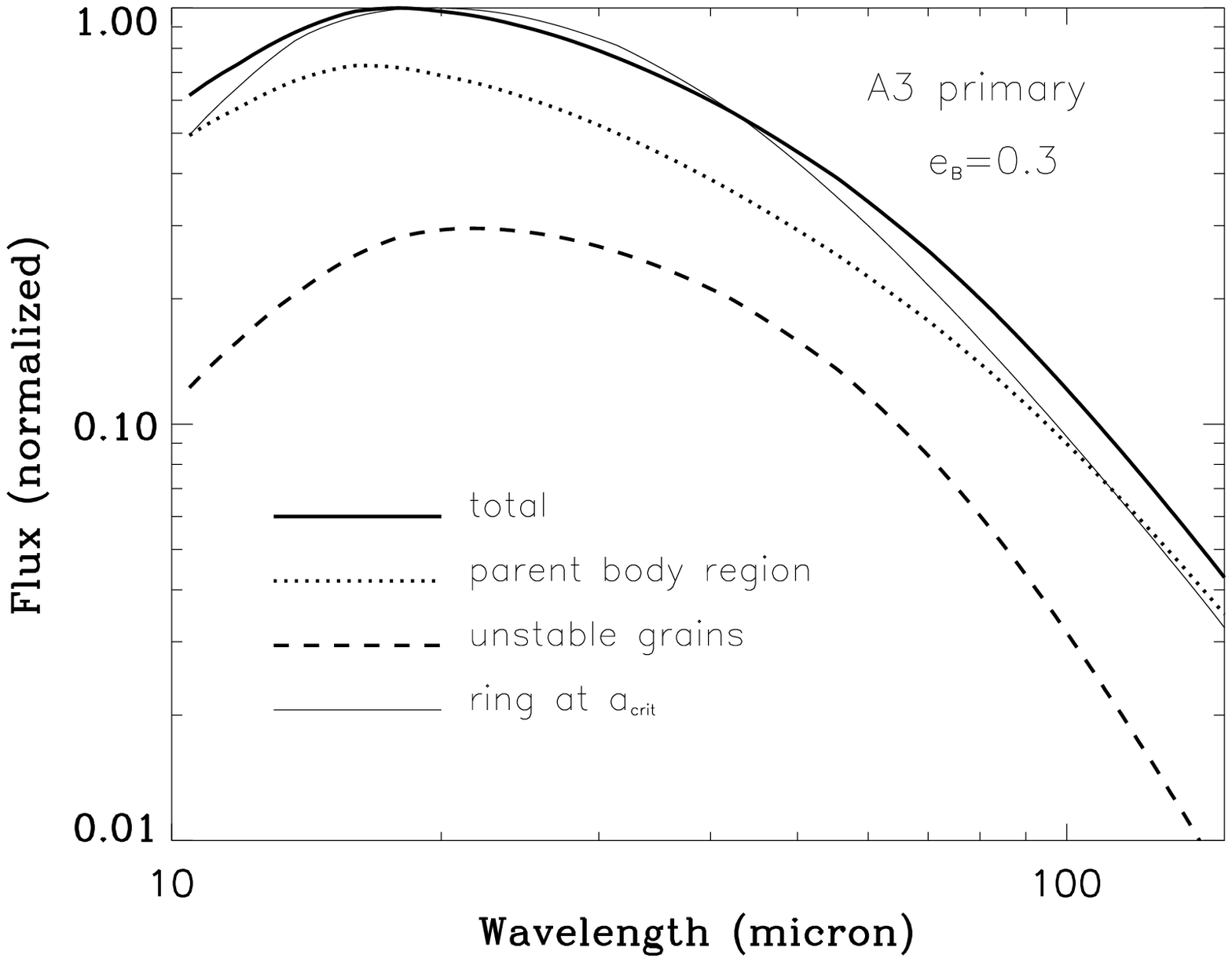}
\hfil
\includegraphics[width=\columnwidth]{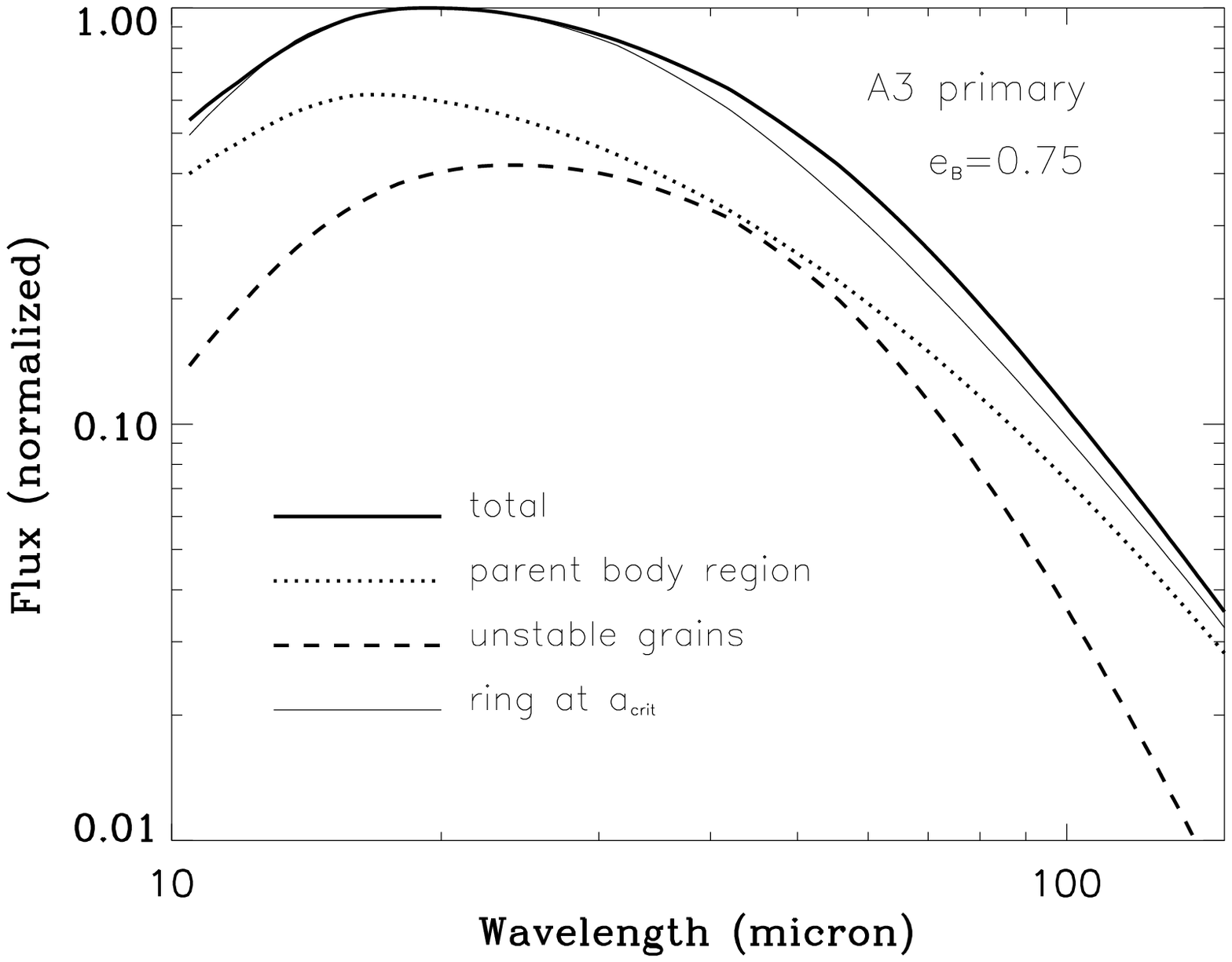}
}
\caption[]{ Synthetic Spectral Energy Distribution (SED) derived for the extended disc case, for two different spectral types for the primary: F5V (upper panel) and A3V (lower panel). For each case, two different $e_b$ values, 0.3 and 0.75, are considered. The companion's semi-major axis is given by the condition that the orbital stability limit is at $a_{crit}=4$\,AU, which leads to $a_b=16\,$AU for the $e_b=0.3$ run and $a_b=50\,$AU for the $e_b=0.75$ one. The respective contributions from $r<a_{crit}$ and $r>a_{crit}$ ("unstable") grains are plotted. For the A3 case, we also plot the SED expected for a reference unperturbed annulus located at $r=a_{crit}$.
}
\label{sed}
\end{figure*}

\subsection{Dust in dynamically unstable regions} \label{unstab}

One important result of our investigation is that a companion star is never able to fully truncate a collisionally active debris disc down to the critical semimajor axis for dynamical stability. More specifically, there is always dust in the dynamically unstable regions of the system, mostly small grains placed on bound eccentric orbits by radiation pressure. Exactly how much material depends on the binary configuration as well as on the disc's optical depth. A faint, less collisionally active disc is less able to fill the $r>a_{crit}$ region than a denser, more massive system (see Fig.\ref{radwide}). Note however that most resolved debris discs have optical depths close to, or even higher than our dense disc case with $\tau =2\times10^{-3}$.

\subsubsection{Signature in thermal emission}

At first glance, this result could help explaining a puzzling result from \citet{tril07}, who detected, for at least 3 binaries, dust in dynamically "forbidden" regions. However, the issue might be more complex than that. The first problem is that, as acknowledged by the authors themselves, for all 3 cases, flux excess is only detected at 70$\mu$m, so that only a maximum dust temperature $T_d$ can be derived (taking the non detection at 24$\mu$m as a lower limit), and thus a minimum distance to the star. There is thus a possibility that the observed dust is further out, possibly on a stable circumbinary orbit (see Fig.5 of that paper). Furthermore, the non-detection at $24\mu$m should preclude the possibility that there is a massive disc closer to the primary, which should in principle be the case because the detected "unstable" dust has to be produced from collisions from more massive parent bodies further in. On a related note, even if multi-wavelengths observations were available, they might not yield a characteristic $T_d$ corresponding to dust beyond $a_{crit}$, because the total thermal flux might remain dominated by grains in the stable parent body region where the optical depth peaks.

We investigate this issue by deriving synthetic Spectral Energy Distributions (SED) from the steady-state optical depth distributions obtained in our dynamical runs. To this effect we need to specify parameters that did not need to be quantified in our numerical study. The first one is the absolute spatial scale, because we need real distances to derive grain temperatures and thermal emission. We consider here a fixed $a_{crit}=$4\,AU value, and explore two cases for the binary eccentricity, $e_b=0.3$ and $e_b=0.75$. This gives a binary separation of 16\,AU in the first case and 50\,AU in the second one. These separations are typical for the intermediate separations binaries for which \citet{tril07} detected grains in the "forbidden" regions. We assume hard sphere silicate grains and calculate their wavelength- and size-dependent optical properties with the Mie theory \citep[optical constants from][]{draine03}. We consider two possible stellar types, F5V and A3V, for which the stellar photospheres are modelled with synthetic NextGen stellar atmosphere spectra for representative stars \citep{hauschildt99}. At any wavelength, the total disc flux, assuming it is unresolved, is obtained by summing up over the distances the thermal emission of all grain sizes weighted by the size distribution. 

For an F5 star (Fig.\ref{sed} A and B), we see that the contribution of unstable grains, although non-negligible (up to 15\% of the flux in the $e_b=0.3$ case and 30\% in the $e_b=0.75$ one), is not enough to drastically modify the total SED. This is mainly because grains in the unstable regions, which have a size $\leq 3\,s_{blow}$, are too small ($s_{blow}\sim 1\mu$m for an F5V star) to significantly contribute at wavelengths $\geq 20\mu$m where, for an F5 central star, their thermal emission should peak if they were perfect emitters. 
For the A3 case (Fig.\ref{sed} C and D), on the contrary, the effect of unstable grains is much more significant. For a companion with $e_b=0.3$, it reaches $\sim 40\%$ in the $20-30\mu$m domain and, more importantly, it is enough to displace the total SED peak emission from 15 to $\sim 22\mu$m. This effect is even more drastic in the $e_b=0.75$ run, where the contribution of unstable grains balances that of $<a_{crit}$ particles for $25\mu$m$\leq \lambda \leq 60\mu$m. The increased importance of $>a_{crit}$ grains for a more massive star is due to the fact that these grains are larger, because $s_{blow} \sim 3\mu$m for an A3V star, and thus better emitters in the mid-IR than for an F5 star. Furthermore, the star being brighter, the disc is hotter and the SED is shifted towards shorter wavelengths, where unstable grains are good emitters. For both $e_b=0.3$ and $e_b=0.75$ cases, the total SED's shift towards longer $\lambda$ is enough for it to peak at a wavelength slightly exceeding that of a ring of particles located at $a_{crit}$. In this case, an analysis similar to that of \citet{tril07} would indeed infer the presence of dust in dynamically forbidden regions \footnote{Although \citet{tril07} used a simplified model, deriving grain characteristic temperatures under the assumption that they behave like black bodies.}. 

Of course, our analysis is simplified. It neglects for instance the fact that grains beyond $a_{crit}$ might not have a temperature corresponding to the radial distance at which they are observed. These grains are indeed on highly eccentric orbits covering a wide radial range and might not have the time to reach thermal equilibrium at a given distance $r$. This issue exceeds the scope of the present study, but our analysis has to the very least shown that grains in the dynamically unstable regions can significantly affect the disc's total SED, possibly making it look as if coming regions beyond $a_{crit}$

\subsubsection{Resolved observations: HR4796} \label{hrfit}

\begin{figure}
\includegraphics[angle=0,origin=br,width=\columnwidth]{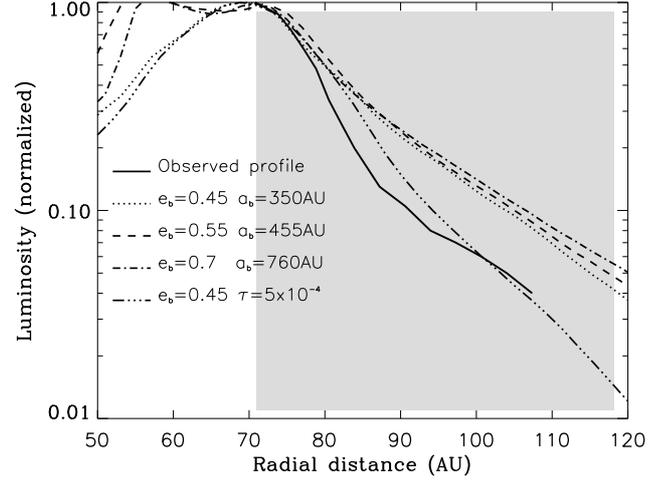}
\caption[]{HR4796A case. Steady state radial luminosity profile obtained for 4 different orbits of the companion and the observationally derived estimate for the disc's optical depth $\tau=7.5\times10^{-3}$, as well as for one alternative $\tau=5\times10^{-4}$ value. The solid line is the scattered light profile observed by \citet{schnei09}.}
\label{hr4796}
\end{figure}

Our results are also applicable to resolved observations in scattered light, as they provide a tool to link the observed spatial structures to the potential influence of a companion star. One direct application are cases epitomized by HR4796, where a companion star, whose orbital parameters are unconstrained, is detected at a large projected distance from a circumprimary disc displaying pronounced features. As an example, HR4796B (at a projected distance $\sim$510AU) has been invoked as a potential source for HR4796A's disc sharp outer edge. \citet{aug99} postulated that if HR4796B's periastron is located at $\sim 200\,$AU, corresponding to $e_{b}\sim 0.4$, then it could efficiently truncate the circumprimary disc at 70AU. However, these estimates were based on simulations by \citet{hall97} of the early truncation of protoplanetary discs that neglected the effect of radiation pressure.

We reinvestigate this scenario more thoroughly to test whether we can get a reasonable fit to the observed scattered light profile of the HR4796a disc with our more complete numerical model. We follow the procedure described in Sec.\ref{methodo} and first run a series of parent body simulations. We find that the minimum eccentricity for a perturber passing at $r\geq 510\,$AU to be able to truncate the parent body disc at $\sim70\,$AU is $e_b = 0.45$. This value is slighly higher than that of \citet{aug99} because we assume the real mass ratio $\mu=0.2$ \citep{jaya98} for the HR4796AB binary instead of the fiducial $\mu=0.5$ taken by \citet{hall97}.
Fig.\ref{hr4796} shows the synthetic radial luminosity profile, in scattered light, obtained for 3 different binary configurations, ranging from $e_b=0.45$ to $e_b=0.7$, and an optical depth $\tau=7.5\times10^{-3}$, an intermediate value between the $\tau \simeq 10^{-2}$ deduced from \citet{aug99} and the $\tau=5\times10^{-3}$ given by \citet{wyatt99}. Scattered light luminosity is derived assuming grey scattering of the grains.

As can be clearly seen, the obtained luminosities in the $r>a_{crit}$ region \footnote{We leave the profiles in the $r<a_{crit}$ region, especially inside the inner edge of the parent body ring at $\sim 60\,$AU, out of our discussion. The sharp $inner$ edge of the HR4796A disc must indeed have a specific origin, unrelated to the companion, which is not the purpose of the present study.} are always much higher than the observed one, even for the lowest possible eccentricity value $e_b=0.45$. 
However, a relatively satisfying fit can be obtained when lowering the disc's optical depth to $\tau=5\times 10^{-4}$. This case might appear to lack physical relevance as it requires an optical depth approximately 15 times lower than the observed one. Nevetheless, this solution can not be directly ruled out because $\tau$ is in our model mostly a way to parameterize the system's collisional activity. Even if there is usually a direct proportionality relation between $\tau$ and $t_{coll}$, there exists some cases where this is not true for the relation between $\tau$ and the collisional $destruction$ timescale $t_{Dcoll}$. This might for instance be the case for dynamically cold discs, where impact velocities are too low to lead to shattering of the impactors, so that $t_{Dcoll}$ would be much higher than $t_{coll}$ deduced from $\tau$. Could HR4796 correspond to such a dynamically cold case? The answer to this question depends on the value of the proper eccentricity $e_p$ of particles within the ring, since $dV_{coll} \sim e_p V_{Kep}$. Assuming a conservative threshold velocity for grain destruction of $50\,$m.s$^{-1}$, this means that, at 70\,AU from the primary, we need $e_p \leq 0.01$. The only observational constraint on $e_P$ is that the ring's width, 14\,AU, gives an upper limit for $e_p$ equal to $\sim 0.1$. A $e_p \leq 0.01$ solution is thus in principle possible, but it might not hold from a theoretical point of view. Debris discs are indeed systems in which the bulk of planetesimal accretion process should already be over and planets or planetary embryos are present and dynamically excite all remaining particles to eccentricities in the $0.05-0.1$ range \citep{arty97,theb07}. As an example, \citet{wyatt99} assumed $\langle e_{p}\rangle=0.13$ for their model of the HR4796A disc.
As a consequence, the low collisional activity solution, even it cannot be fully dismissed, does not appear to be the most generic one. Furthermore, even $if$ the disc is dynamically cold, then a sharp outer edge might be maintained $without$ the need for an external perturber, as low $e_p$ values might naturally create a depletion of small high-$\beta$ grains \citep[see][]{theb08}.
The companion is thus probably not a satisfying explanation for the HR4796A disc's radial profile.

Our conclusion differs from that of \citet{aug99} because \citet{hall97} did not consider the effect of radiation pressure, implicitly considering particles large enough to be only affected by gravity. This would correspond to the "parent body" disc of our study, for which we indeed find that it is truncated by companion stars having similar characteristics as in \citet{hall97}. But the present numerical exploration has shown that, even if HR4796B is eccentric enough to cut-off the circumprimary disc of parent bodies, it cannot do the same job for the small debris steadily produced by collisions amongst these parent bodies. 

More generally, these results might be useful for all systems where a bound, or possibly bound companion is detected far away from a circumprimary debris disc and is considered as a potential source for pronounced features in the disc, especially sharp edges, under the assumption that it is on a highly eccentric orbit allowing it to pass in the disc's vicinity. The present numerical exploration shows that the effect of such a companion should in fact be limited, because companions on eccentric orbits are less efficient in preventing matter from populating dynamically unstable region.

\subsection{Small grain depletion in dynamically stable regions} \label{stab}

In apparent contrast to the previous results showing the reduced effect of companion stars in the dynamically unstable regions, we find that binarity can also significantly affect a debris disc in the dynamically {\it stable} region, depleting it from a large fraction of its small grain population. This is once again due to radiation pressure: small particles produced at $a<a_{crit}$ spend a large fraction of their eccentric orbits beyond $a_{crit}$, where they can be removed by companion star perturbations. As a net result, small grains with sizes smaller than $\simeq 3s_{blow}$ (radiation pressure blow out size), will be underabundant in the collisionally active parent body region. This result is significant even if the parent body region lies well inside $a_{crit}$, as shown in our inner ring case displayed in Fig.\ref{betna}.

Can this lack of very small grains have an observable counterpart? {\it A priori}, this should be the case for wavelengths comparable to a given system's $s_{blow}$. Indeed, in the normal unperturbed case, these discs' total cross section should be dominated by the smallest bound grains, close to $s_{blow}$ \citep[see for example][]{theb07}, so that the disc should appear red for multi-wavelengths observations around $\lambda \simeq s_{blow}$. This seems for instance to be the case for HR4796A \citep{schnei09}, and might thus be an additional argument to rule out the binary sculpting scenario. Conversely, in the absence of $s_{blow}<s<3\,s_{blow}$ grains, the disc should appear "grey" with no clear colour trend.

Nevertheless, even if this effect is in principle qualitatively observable, these results should be taken with caution. Firstly because the size range for grain depletion is relatively limited (a factor 3), so that precise and well sampled (in wavelengths) observations should be necessary. In addition, different grain composition and porosity might greatly complicate this picture, changing the dynamical behaviour of a particle of a given size. Finally, there might be other mechanisms that can lead to a depletion of small grains \citep[as for instance the dynamically cold case in][]{theb08}.

\section{Summary and Conclusions}

We have numerically explored the response of a circumprimary debris disc to the presence of a stellar companion. Our main results can be summarized as follows:

\begin{itemize}
\item A companion star is never able to fully truncate a a collisionally active circumprimary disc. Even if its perturbations place a critical distance $a_{crit}$ to the primary beyond which orbits are unstable, the $r>a_{crit}$ regions are not empty. They are populated by small grains, steadily produced by collisions from parent bodies in the stable regions, which are placed on eccentric orbits by radiation pressure
\item The amount of material in the unstable regions depends on the balance between the rate at which small grains are produced by collisions and the rate at which they are removed by the companion's perturbations. Massive discs with short collisional timescales are thus the most favourable cases, for which the system's total cross section can become dominated by the unstable grains. Conversely, for a fixed $a_{crit}$, binaries with circular orbits are more efficient in removing $r>a_{crit}$ grains than binaries on highly eccentric orbits. 
\item For massive, A-type primary stars, the contribution of the unstable grains can significantly affect the disc's SED. For dense discs and/or eccentric binaries, the global SED can peak at a wavelength longer than that corresponsping to grains located at $a_{crit}$.
\item For discs that are resolved in scattered light and display sharp outer edges, our study shows that it is difficult to explain these outer truncations by a distant companion star on an eccentric orbit. For the specific case of HR4796A, we show that the disc is probably too massive (i.e., collisionally active) and the detected companion on a too-eccentric orbit to explain the observed structure of the circumprimary disc.
\item Further putting into perspective the role of $a_{crit}$ as an absolute frontier is the fact that the companion star is able to affect the grain population within the stable region. The $r<a_{crit}$ region is depleted from grains within a factor 2-3 of the blow out size by radiation pressure. However, this absence of small particles might not be easily observable.
\end{itemize} 

\begin{acknowledgements}
PT wishes to thank Mark Wyatt for enlightening discussions. The authors thank the anonymous reviewer for very useful comments
\end{acknowledgements}

{}
\clearpage

\end{document}